\documentclass[aip,floatfix,amsmath,amssymb,reprint]{revtex4-1}
\pdfoutput=1
\usepackage{graphicx}
\usepackage{dcolumn}
\usepackage{bm}

\usepackage[utf8]{inputenc}
\usepackage[T1]{fontenc}
\usepackage{mathptmx}
\usepackage{etoolbox}
\usepackage{subfigure}
\usepackage{color,soul}
\usepackage{comment}
\usepackage{orcidlink}
\usepackage{placeins}

\draft 

\begin{document}


\title{Counter-Streaming Beams in Collisionless Pair Plasma Instability Systems III: Collisionless Heating, Acceleration, and Radiation} 



\author{Michael C. Sitarz \orcidlink{0000-0001-9003-0737}}
\affiliation{Department of Physics and Astronomy, University of Kansas, Lawrence, KS 66045}
\email{mcsitarz@ku.edu}

\date{April 23, 2024}

\begin{abstract}
Energetic astrophysical phenomena, such as $\gamma$-ray bursts and supernova explosion-driven shocks in collisionless plasmas, involve various plasma kinetic instabilities, such as the Weibel instability. These systems support various types of particle acceleration and radiation through a variety of mechanisms. In this paper, we explore the energy transformations and dynamical effects of violent filament mergers seen between the large current filaments formed by the Weibel instability. The radiative processes involved in the Weibel instability filament building are also discussed in relation to jitter and synchrotron radiation.
\end{abstract}

\pacs{}

\maketitle 

\section{Introduction}
\indent Collisionless discontinuities in the form of shock waves are a common phenomenon arising from supernova blast waves (SN) \cite{1}, quasar jets \cite{2,3}, solar flares \cite{4}, pulsar wind nebula (PWN) \cite{5}, relativistic jets in active galactic nuclei (AGN), and gamma ray bursts (GRB) \cite{6}. All of these high energy events show signs of particle acceleration along with the production of magnetic fields. Weibel \cite{7} and Fried \cite{8} demonstrated that magnetic fields arise from an instability driven by an anisotropy in the particle distribution's temperature or velocity space. This anisotropy brings forth a plasma instability responsible for runaway magnetic fields. One common set up for this system is used in this series of papers - counter streaming beams. These beams are composed of a $e^\pm$ pair plasma, no ions or protons present. Focus is often spent on the radiation generated from this system (for example, Jitter radiation produced as a by-product of particle acceleration during filament building \cite{9}) and the various synchrotron regimes the system can produce. There is also question if this environment can play host to particle excitation and collisionless heating. This paper puts forth an analysis focused on the the heating and excitation of particles in a collisionless pair plasma system. \\

\indent The first paper in this series, \cite{MCS_WP1}, hereafter referred to as Paper I, studied the parameterization effects of the Weibel instability (WI) and its subsequent evolution. Paper I also explored the spectral signals in the dispersion relations, identifying the dominate isotropic electromagnetic mode and showed that there exists a secondary (and later dominate) signal in the beam parallel $E_x$ belonging to the two stream instability (TSI). The second paper, \cite{MCS_WP2}, hereafter referred to as Paper II, analyzed the discrete spectral signals of the simulations using the spectral cone technique. This analysis brought forth the conclusions that the spectral waves interact with the physical filaments of the system and therefore could accelerate particles. This acceleration was seen in the spectral power evolution as energy dipped and peaked along the physical field evolution.  \\

\indent This paper presents a study on the environment seen in Papers I and II thought to host particle acceleration. This highly magnetic and turbulent local environment is thought to be ideal for collisionless heating phenomenon brought on by violent filament mergers seen in the control simulation (S1) from the previous two studies. Energy transformations within a heating environment are thought to accelerate (or decelerate) particles, confirming the previously mentioned theories. Radiation environments from the spectral scale of the magnetic field are also shown, with implications of Jitter and synchrotron radiation environments and their convalescence also discussed. \\

\indent The remainder of the paper will be organized as follows: \S$2$ is a brief  review of the Weibel instability, \S$3$ is a review of the Two Stream instability, \S$4$ introduces collisionless heating and the justification of its existence within the system along with radiative discussions. \S$5$ presents the results of the study and \S$6$ will contain concluding remarks and possible implications.\\

\section{Weibel Instability}
\subsection{Weibel Instability Timeline}
\indent The Weibel instability (WI) is proposed to be the source of intense magnetic fields within the GRB prompt emission, afterglow \cite{9} and astrophysical shock frames \cite{11}. These claims were later proved numerically by a number of sources \cite{12,13,14,15,16}. The magnetic field turbulence generated by the WI operating at the shock front \cite{9} is of a few ion skin depths \cite{17}, is sub-Larmor in scale \cite{18}, and is an important mechanism in the mediation of the GRB shock \cite{19}. This small scale turbulence is not uncommon in an astrophysical setting, with other examples being the electromagnetic Whistler, filamentation, and mixed modes or electrostatic (ES) Langmuir oscillations \cite{9}. \\

\indent The WI was derived by Weibel in $1959$ using a fully kinetic analysis \cite{7}, where he considered a non-relativistic plasma with an anisotropic particle distribution function (PDF). It begins as a very weakly or non-magnetized plasma with an anisotropic velocity distribution of electrons and fixed ions, with the electron temperature dependency based on direction (bi-Maxwellian anisotropy). 
\begin{equation}\label{Eq:Bimax_Dist}
    f(v) = \frac{n}{\sqrt{\pi^3}\Theta^2_\perp\Theta^2_\parallel}exp\left[-\left(\frac{v^2_\perp}{\Theta^2_\perp} + \frac{v^2_\parallel}{\Theta_\perp^2}\right)\right],
\end{equation}
where the $\Theta$ terms are defined as (noting that $T_\perp \ne T_\parallel$).
\begin{equation}\label{Eq:Theta_Temp}
    \Theta_{\parallel / \perp} = \sqrt{\frac{2kT_{\parallel / \perp}}{m}}.
\end{equation}
The total distribution for a particle species $s$ can be given as the sum of equilibrium and perturbed positions 
\begin{equation}\label{Eq:Total_Dis_Species}
    f_s = f_{0s} + \hat{f}_s.
\end{equation}
This then generates a transverse mode of perturbations in the electrons.\\

\begin{figure}[h]
    \centering
    \includegraphics[scale=0.1]{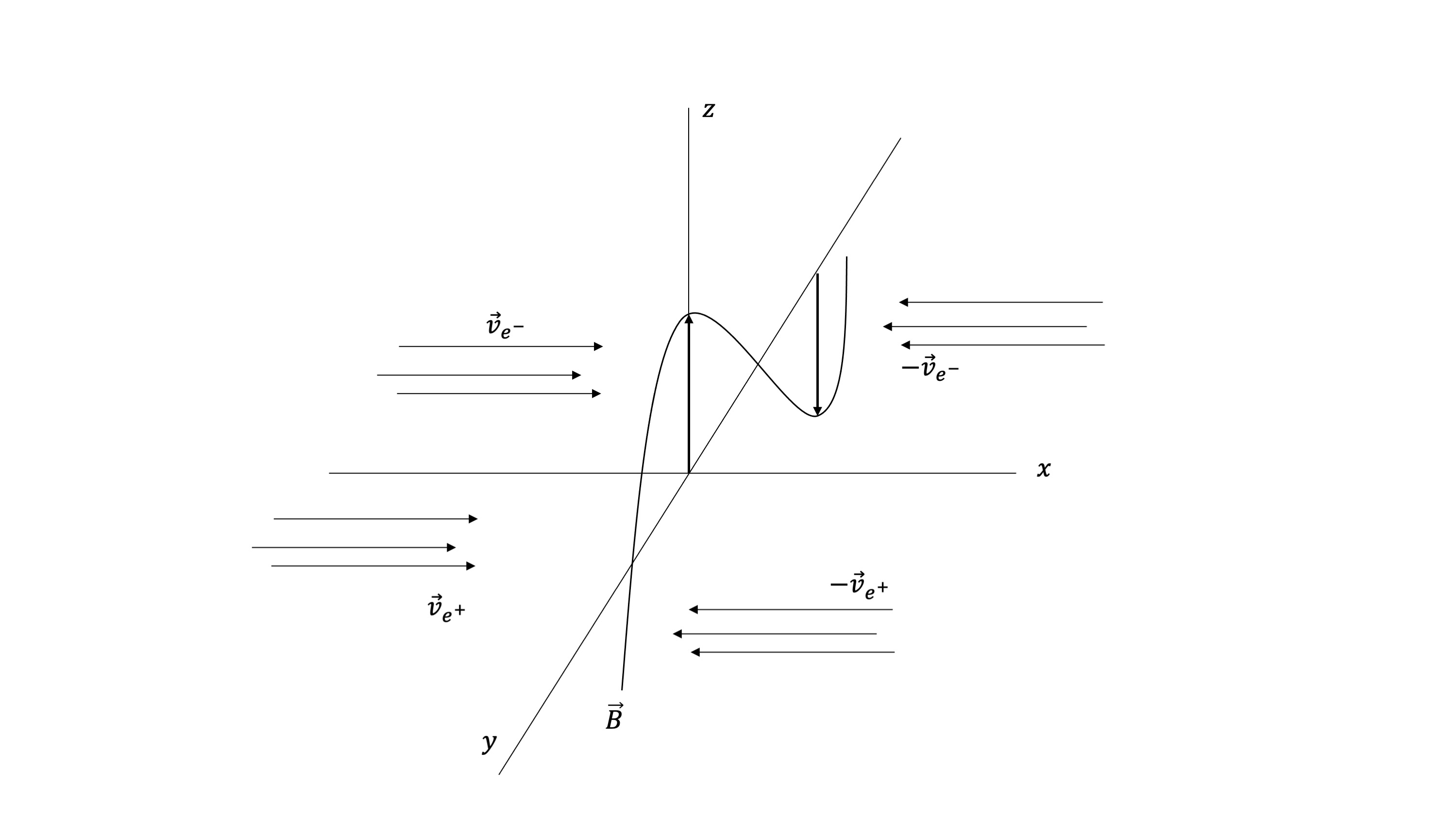}
    \caption{Example set up of two counter streaming beams with a sinusoidal mode $\Vec{B}$ field normal to the streaming plane. The beams are composed of both positrons and electrons with $n_{e^-} = n_{e^+}$.}
    \label{Fig:Weibel_Cartoon}
\end{figure}

\begin{figure}[h]
    \centering
    \includegraphics[scale=0.1]{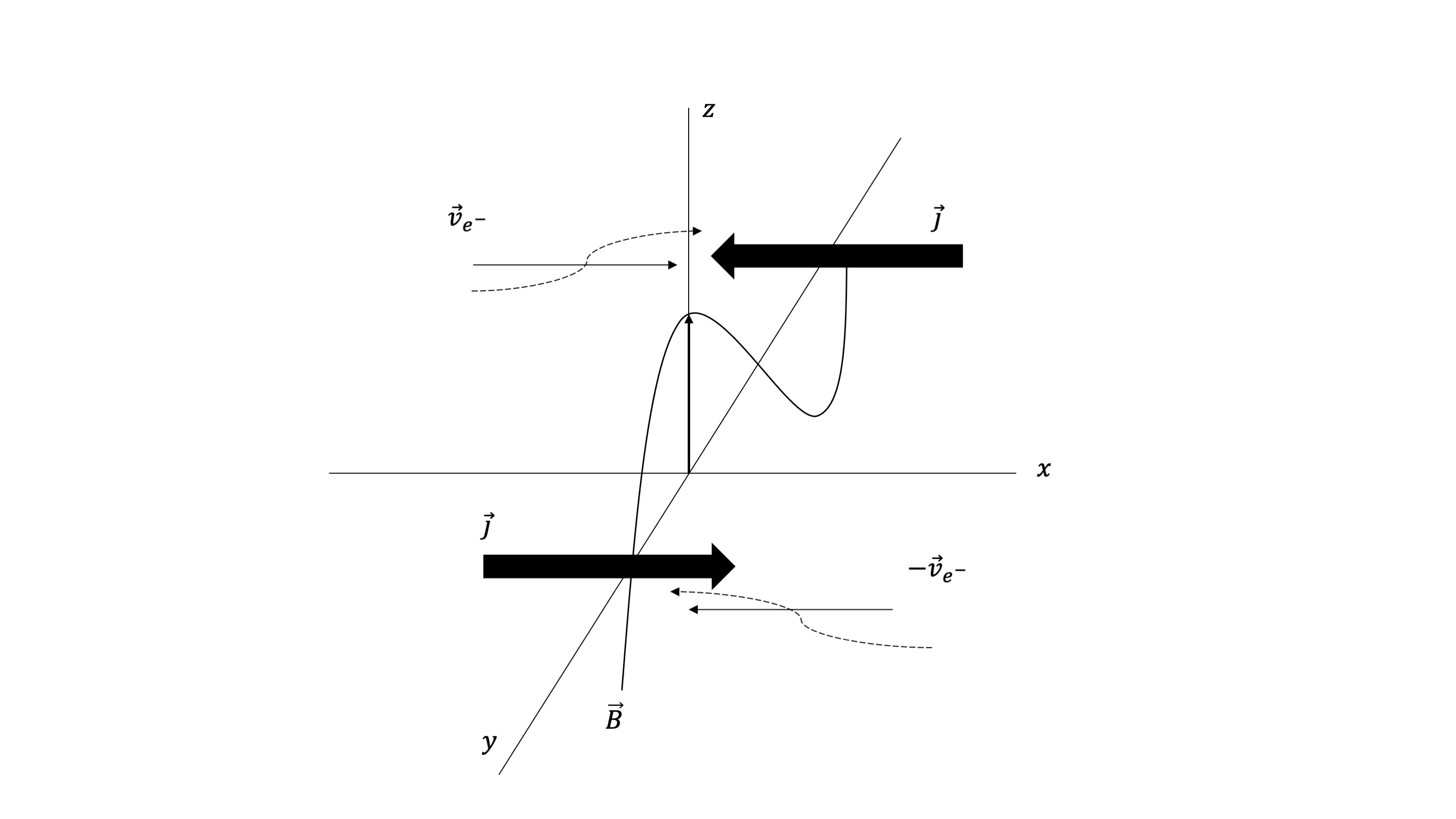}
    \caption{As the beams stream into the $\Vec{B}$ field, the particles' trajectory begins ti deflect towards the nodes of the field. These particles bunches then produce the signature current filaments in the direction of flow.}
    \label{Fig:Filament_Cartoon}
\end{figure}

\indent Fried \cite{8} treated a anisotropic PDF more generally as a two stream in a cold plasma. He describes counter streaming planes of electrons with a very small sinusoidal magnetic field normal to the streaming plane, producing a runaway effect (Fig. \ref{Fig:Weibel_Cartoon}). This sinusoidal field is now treated as initial fluctuations in the system.
The Lorentz force ($\vec{F} = \frac{e}{c}(\vec{v} \times \vec{B})$) acts on the charged particles and deflects their trajectories. This results in spatial concentrations of particles that generate separate current filaments. $\vec{B}$ in the filaments increases the initial fluctuation (Fig. \ref{Fig:Filament_Cartoon}). 

\subsection{Current Filament Theory}
\indent It is important to discuss the dynamics and evolution of the current filaments present in the WI system, as they are the most salient features of the instability process. This discussion is derived from \cite{20} and the sources therein. At the beginning of the instability process, as particles begin to deflect and bunch, small filaments form within the box. This example will consider one-dimensional filaments, but the theory is identical in higher dimensions without loss of generality. All filaments start out identical in size and quantity (charge and current neutrality must be preserved). They possess initial characteristics as follows: diameter $D_0$, mass per unit length $\mu_0 \simeq 0.25mnD_0^2\pi$ (particle mass $m$ and number density $n$), current $I_0$, and spatial separation $d_0 \simeq 2D_0$ from center to center. They form randomly distributed throughout the box and at rest but vulnerable to attraction/repulsion forces. \\

\indent The spatial distribution of filaments appears as as a regular pattern post saturation, but they are not stationary. The equation of motion for the filaments can be easily found. The magnetic fields produced by a filament is given by 
\begin{equation}\label{Eq:Filament_Mag_Field}
    B_0(r) = \frac{2I_0}{cr},
\end{equation}
with cylindrical radius $r$. Assuming the filament is straight (no kinks or bends), the force per unit length is given as
\begin{equation}\label{Eq:Filament_Force_Length}
    \frac{dF}{dl} = \frac{-B_0I_0}{c}.
\end{equation}
Allowing
\begin{equation}\label{Eq:Filament_EoM}
    \Ddot{x} = \frac{-2I_0^2}{c^2\mu x},
\end{equation}
with $r = 2x$ and reduced mass $\mu_r = \mu_0/2$, the force per unit length can then be represented by
\begin{equation}\label{Eq:Fil_Force_Length_Reduced}
    \frac{dF}{dl} = \mu \Ddot{x}.
\end{equation}

\indent With their movement modeled, the merger time can now be estimated. Merger is first defined as when two filaments touch, as opposed to a full integration. This occurs at $d_0 \simeq D_0$. The time scale of filament mergers is independent of the merger process itself, as the coalescence of two current filaments involves complex current redistribution dynamics. The force interaction between filaments gets weaker as distance increases, limiting merger rate. The time scale $\tau$ can be readily estimated from $\Ddot{x}$ given that $\Ddot{x} \sim (d_0/2)/\tau_0^2$. For non-relativistic motion
\begin{equation}\label{Eq:NR_Fil_Merge_Time}
    \tau_{0,NR} \sim \sqrt{\frac{D_0^2c^2\mu_0}{2I_0^2}},
\end{equation}
where the max velocity of a filament is limited to 
\begin{equation}\label{Eq:Fil_Max_Velo}
    \vec{v}_{Max,0} \sim \frac{D_0}{2\tau_0} \sim \frac{I_0}{\sqrt{2c^2\mu_0}}.
\end{equation}
If filament motion becomes comparable to $c$ ($t(x) \simeq \frac{x}{c}$) then the regime becomes relativistic and the time scale now reads
\begin{equation}\label{Eq:R_Fil_Merge_Time}
    \tau_{0,R} \simeq \frac{d_0c}{2} = \frac{D_0}{c}.
\end{equation}

\indent Filament merging is a hierarchical and self similar process. If the system initially has $N_0$ number of filaments, filaments will merge pairwise under the given time scales into a new ($1^{st}$) generation of filaments numbering $N_0/2$. The properties of the initial filaments scale as the $k^{th}$ generation is produced, following
\begin{multline}\label{Eq:Number_of_kth_Gen_Fil}
    I_k = 2^kI_0, \ \mu_k = 2^k\mu_0, \ D_k = 2^{k/2}D_0, \\
    d_k \sim \frac{D_k}{2}, \ \tau_{k,NR} = \tau_{0,NR}, \\
    \tau_{k,R} = 2^{k/2}\tau_{0,R}.
\end{multline}

\indent Filament motion is not constrained in the non-relativistic or relativistic regimes. Velocities will increase with each generation merged, following $\vec{v}_{Max,k} = 2^{k,0}\vec{v}_{Max,0}$. Non-relativistic can become relativistic following  $j = 2log_2(c/\vec{v}_{Max,0})$ mergers. \\

\indent Finally, filament parameters and correlation scales can be expressed as a function of physical time. Based on merger levels, it will take $t = \sum^k_{k'=0} \tau_{k'}$ to complete the $k$ mergers. Using this relation, merger level can be shown as
\begin{equation}\label{Eq:Fil_Merge_Level}
    k \simeq 2log_2(\frac{t}{\tau{0,R}}),
\end{equation}
with characteristic magnetic field length
\begin{equation}\label{Eq:Char_Fil_Mag_Length}
    \lambda_B(t) = D_02^{t/\tau_{0,NR}}.
\end{equation}

\section{Two Stream Instability}
\subsection{Instability Origins}
\indent The two stream instability (TSI) is one of the most ubiquitous instabilities found in plasma physics. It occurs when two species of particles have counter propagating drift velocities $v_0$. From this basic physical set-up, further system parameters can dictate the unstable electrostatic modes within the dispersion relation. \\

\indent There are two general beam cases when considering the TSI, cold and hot, that both saturate when the beam particles are bound within the electric field of the propagating wave. For hot beams, TSI can be thought of as a type of inverse Landau damping. There is a small population of particles with a drift velocity greater than that of the phase velocity of the propagating wave. The majority of particles are slower than the phase velocity and there are particles with equal velocities to the wave. Regarding the instability system itself, when a hot beam of electrons is injected into a stationary background, the velocity space distribution is said to posses a ``bump on tail'' distribution function. On this function, if the phase velocity of the excited wave exists in a region of positive slope (more particles faster than its phase velocity than slower) there exists a greater energy transfer from the fast particles to the slower wave, further exciting the wave. For cold beams, none of the particles in either beam possess a drift velocity equal to that of the phase velocity of the wave within the system (resonance). The beam particles are clustered in physical space in a propagating wave. This motion becomes self-reinforcing despite no resonance. The excitation of a TSI in the cold beam limit is in Paper II of the series \citep{MCS_WP2}.\\

\section{Heating and Radiation}
\subsection{Collisionless Heating}
\indent While heating is though of as a by product of collisions within the plasma (Ohmic heating), it can and has been shown to occur in collisionless plasma systems with an electron-ion composition in Weibel-mediated shocks \citep{VLG2022}. It has also been shown that Alfevenic turbulence injects energy in the form of heat-energy, again in an electron-ion system \citep{KBS2018}. This work continues its focus in the Weibel Instability system with a $e^\pm$ pair-plasma and uses a different approach to show collisionless heating.\\

\indent Heating in a collisionless system may come in the form of Joule heating. This is seen directly from the Poynting Theorem \citep{Poynting}. Considering a distribution of particles in the system under the influence of electromagnetic fields, the work done on the charge can be described by the Lorentz force
\begin{equation}
    \frac{dW}{dt} = q\vec{v} \cdot \vec{E}.
\end{equation}
Given a volume $V$ (simulation box) and a distribution of charges, the above now states
\begin{equation}
    \frac{dW}{dt} = \int_V \vec{J} \cdot \vec{E} d^3x.
\end{equation}
Recalling Ampere's Law $\vec{J} = \nabla \times \vec{B} - \partial\vec{E}/\partial t$ and substituting and distributing it into our expression gives
\begin{equation}
     \int_V \vec{J} \cdot \vec{E} d^3x = \int_V d^3x \left(\nabla \cdot (\vec{E} \times \vec{B}) - \vec{B} \cdot (\nabla \times \vec{E}) + \vec{E} \cdot \frac{\partial \vec{E}}{\partial t} \right),
\end{equation}
here, the vector identity of the divergence of a cross product was used. Here it is noted that the traditional, macroscopic fields of $\vec{D}$ and $\vec{H}$ are replaced by the microscopic fields of $\vec{E}$ and $\vec{B}$. This is further explained below. We now introduce a second Maxwell equation, Faraday's Law $\nabla \cdot \vec{E} = -\partial \vec{B}/\partial t$ into the second term
\begin{equation}\label{Eq:FullExpansPoy}
     \int_V \vec{J} \cdot \vec{E} d^3x = \int_V d^3x \left(\nabla \cdot (\vec{E} \times \vec{B}) + \vec{B} \cdot \frac{\partial \vec{B}}{\partial t} + \vec{E} \cdot \frac{\partial \vec{E}}{\partial t} \right).
\end{equation}

\indent We pause the derivation here for careful treatment of the next assumptions. First, the total electromagnetic energy density is traditionally defined by 
\begin{equation}
    u = \frac{1}{2}(\vec{E} \cdot \vec{D} + \vec{B} \cdot \vec{H}).
\end{equation} 
To begin, $\vec{D} = \vec{E} + \vec{P}$ represents the displacement field as a sum of the electric and polarization fields (unit bearing constants have been left out). Defining $\vec{P} = \chi_e\vec{E}$ with susceptibility $\chi$ in a traditional way does little to illuminate its purpose here, if any. Redefining $\vec{P}$ into a polarization density $d\vec{p}/dV$ is clearer. The polarization density is a result of bound charge density $\rho_b$. The nature of the system (pair plasma) allows the particles to be described as free charges, not bound. This, in turn, renders the polarization to be zero and recover $\vec{D} = \vec{B}$. The same assumptions and logic can be applied to the auxiliary magnetic field $\vec{H}$. Defined by $\vec{H} = \vec{B} - \vec{M}$, where $\vec{M}$ is the magnetization. In the same vein as polarization, magnetization (or its density) describes the density of permanently induced dipoles. Again, our system does not contain or create and permanence in the a particles or their features, rendering $\vec{M} = 0$ and $\vec{H} = \vec{B}$. With only $\vec{E}$ and $\vec{B}$ present, we look to model only effects of the electrical current. The energy density equation now reads
\begin{equation}
    u = \frac{1}{2}(\vec{E} \cdot \vec{E} + \vec{B} \cdot \vec{B}).
\end{equation}

\indent From the redefined energy density, we recall the vector calculation for some vector $\vec{A}$
\begin{equation}
    \frac{\partial}{\partial t}(\vec{A} \cdot \vec{A}) = \frac{\partial}{\partial t}A^2 = 2A\frac{\partial}{\partial t}A = 2\vec{A} \cdot \frac{\partial}{\partial t}\vec{A},
\end{equation}
which is seen in the second and third terms on the right hand side (RHS) of (Eq. \ref{Eq:FullExpansPoy}). Substituting recovers
\begin{equation}
    \int_V \vec{J} \cdot \vec{E} d^3x = \int_V d^3x \left(\nabla \cdot (\vec{E} \times \vec{B}) + \frac{\partial u}{\partial t} \right).
\end{equation}
Inserting the definition of the Poynting vector and algebraically rearranging recovers the familiar expression for Poynting's theorem
\begin{equation}
    \frac{-\partial u}{\partial t} = \nabla \cdot \vec{S} + \vec{J} \cdot \vec{E},
\end{equation}
where $\vec{S} = \vec{E} \times \vec{B}$ is the Poynting vector (studied w.r.t this system in Paper II \citep{MCS_WP2}) modeling the energy flow out of the system and $\vec{J} \cdot \vec{E}$ is the rate at which the fields do work in the system. Here, $\vec{J}$ is the current density of the motional charges. This can be considered a mechanical power, as opposed to radiative. The full theorem details that the difference of electromagnetic energy entering and leaving a system must either dissipate or convert into a different form of energy. This conversion is often to heat. \\

\indent The second careful assumption comes in the form of dispersivity of media. A standard treatment assumes the charges are moving through a medium which is not dispersive. We may ignore this requirement based on two assumptions: the original statement refers non-negligible macroscopic effects and imposed scale in-variance. We have shown that our system does not work in macroscopic regimes, eliminating macroscopic effects and their associated fields $\vec{D}$ and $\vec{H}$. In macroscopic media, EM effects are models as the spatially averaged displacement and auxiliary fields. The Poynting theorem in large, macroscopic systems can be viewed as a self-consistently defined microscopic system, therefore retaining the validity\citep{SPoy2009}. A macroscopic system can be defined as various, local microscopic environments while retaining its formalism predicted by the full theory. It is this scale resolving feature that allows us to use the Poynting theorem as a model for collisionless heating in a low-loss system.\\

\indent This simple but powerful formulation of energy convergence to heat is integral in the next steps of this study. If energy is converted to heat in a region, energy is also reasonably converted into particle motion/acceleration. The next steps of this study are to confirm and validate singular particle motion in an effort to uncover the radiation of the system.\\

\subsection{Relativistic Motion} 
\indent Development of further theorises requires the equations of motion of relativistic particles. While this study does not analyze individual particles or their individual signatures, understanding the kinematics is crucial to the remainder of the work. What follows is a sample derivation that can be commonly found in many texts.\\

\indent Begin with Lorentz Force
\begin{equation}\label{Eq:Lorentz_Force}
    \frac{dP^{\mu}}{d\tau} = \frac{q^\pm}{c}F^\mu_\nu U^\nu,
\end{equation}
where
\begin{equation}\label{Eq:F_mu_nu}
    F^\mu_\nu = 
        \begin{bmatrix}
        0 & E_x & E_y &  E_z \\
        E_x & 0 & B_z & -B_y \\
        E_y & -B_z & 0 & B_x \\
        E_z & B_y & -B_x & 0 \\
        \end{bmatrix}.
\end{equation}
Replacing the relativistic momenta $P^\mu$ with is complete representation, where Lorentz factor $\gamma$ will be referred to as $\gamma_{B}$ for beam velocity and $v_i$ will be the particle's velocity. $U^\nu$ is replaced with the velocity matrix and beam velocity.
\begin{equation}\label{Eq:Matrix_Lorentz_Force}
    \frac{d}{d\tau}\gamma_{B}m_e\begin{bmatrix}
        c \\
        v_x \\
        v_y \\
        v_z \\
    \end{bmatrix}
    = \frac{q^\pm\gamma_{B}}{c}\begin{bmatrix}
        0 & E_x & E_y &  E_z \\
        E_x & 0 & B_z & -B_y \\
        E_y & -B_z & 0 & B_x \\
        E_z & B_y & -B_x & 0 \\
        \end{bmatrix} \begin{bmatrix}
        c \\
        v_x \\
        v_y \\
        v_z \\
    \end{bmatrix}.
\end{equation}
Replacing proper time $\tau$ with $dt/\gamma_B$ and multiplying matrices recovers
\begin{equation}\label{Eq:NonRelTime_Lorentz_Force}
    \frac{d}{dt}\gamma_B^2m_e\begin{bmatrix}
        c \\
        v_x \\
        v_y \\
        v_z \\
    \end{bmatrix} = \frac{q^\pm\gamma_{B}}{c}\begin{bmatrix}
        E_xv_x + E_yv_y + E_zv_z \\
        E_xc + B_zv_y - B_yv_z \\
        E_yc - B_zv_x + B_xv_z \\
        E_zc - B_yv_x - B_xv_y \\
        \end{bmatrix}.
\end{equation}
Before reducing the dimensions to our system, we take note that this can be simplified to a more recognizable form
\begin{equation}\label{Eq:Simplified_Lorentz_Force}
    \frac{d}{dt}\gamma_B m_e \begin{bmatrix}
        c \\
        \vec{v}
    \end{bmatrix} = \frac{q^\pm}{c}\begin{bmatrix}
        \vec{E} \cdot \vec{v} \\
        \vec{E}c + \vec{v} \times \vec{B}
        \end{bmatrix},
\end{equation}
with the two equations within the matrix representing the energy
\begin{equation}\label{Eq:Energy_Eq}
    \frac{d}{dt}\gamma_Bm_ec^2 = q^\pm \vec{E} \cdot \vec{v},
\end{equation}
and the momentum
\begin{equation}\label{Eq:Momentum_Eq}
    \frac{d}{dt}\gamma_Bm_e\vec{v} = q^\pm(\vec{E}c + \vec{v} \times \vec{B}).
\end{equation}
It is worthwhile to note that if $\vec{E} = 0$, then radiation losses can be ignored via $\frac{d}{dt}\gamma_Bm_ec^2 = 0$. Because the system does not have a zero electric field, we can safely say we have non-constant velocity.\\

\indent Returning to Eq. (\ref{Eq:NonRelTime_Lorentz_Force}), the expression can be simplified by noting $v_z = E_z = B_x = B_y = 0$
\begin{equation}\label{Eq:Reduced_Lorentz_Force}
    \frac{d}{dt}\gamma_B^2m_e\begin{bmatrix}
        c \\
        v_x \\
        v_y \\
    \end{bmatrix} = \frac{q^\pm\gamma_{B}}{c}\begin{bmatrix}
        E_xv_x + E_yv_y \\
        E_xc + B_zv_y \\
        E_yc - B_zv_x \\
        \end{bmatrix}.
\end{equation}
This recovers three coupled equations,
\begin{equation}\label{Eq:2D_Energy_Eq}
    \frac{d}{dt}\frac{\gamma_Bm_ec^2}{q^\pm} = E_xv_x + E_yv_y,
\end{equation}
\begin{equation}\label{Eq:2D_x_Momentum_Eq}
    \frac{d}{dt}\frac{\gamma_Bm_ev_x}{q^\pm} = E_xc + B_zv_y,
\end{equation}
\begin{equation}\label{Eq:2D_y_Momentum_Eq}
    \frac{d}{dt}\frac{\gamma_Bm_ev_y}{q^\pm} = E_yc - B_zv_x.
\end{equation}
It is worthwhile to note that $\vec{v_{\parallel}}$ in this frame (parallel to the magnetic field) is $\vec{v_z} = 0$ and $\vec{v_\perp} = \sqrt{\vec{v_x}^2 + \vec{v_y}^2} \neq 0$. All parallel movement and its derivatives are constant in these calculations ($0$). Moving constants algebraically, we are left with the final form of our equations,
\begin{equation}\label{Eq:2D_Energy_Eq_Final}
    \frac{d\gamma_B}{dt} = \frac{q^\pm}{m_ec^2}(E_xv_x + E_yv_y),
\end{equation}
\begin{equation}\label{Eq:2D_x_Momentum_Eq_Final}
    \frac{d\gamma_B v_x}{dt} = \frac{q^\pm}{m_ec}(E_xc + B_zv_y),
\end{equation}
\begin{equation}\label{Eq:2D_y_Momentum_Eq_Final}
    \frac{d\gamma_B v_y}{dt}= \frac{q^\pm}{m_ec}(E_yc - B_zv_x).
\end{equation}

\subsection{Radiation}
\indent Particles traveling at relativistic speeds in curved paths (tangential acceleration) emit synchrotron radiation. In this system, a curved path can be described as $\delta v_i > \delta v_j$. In the current numerical setup, the beams propagate along the $x$-direction. For the discussion for early-system synchrotron radiation, we may disregard any $\delta v_x$ due to $v_x >> v_y$ and therefore $\delta v_x <<< \delta v_y$. The deflection of particles from the beam into filament nodes is an example of curvature we will see in the early stages of the system. Later on, during filament mergers, no restrictions are applied to the particle radiation. The following derivation is based off of one those found in \citep{RL1979}. \\

\indent We begin with non-relativistic particles to set up our formalism. This skips the derivation of the Lienard - Wiechart Potentials and subsequent radiation fields, which can be found in many electromagnetism textbooks. Starting with general radiation fields
\begin{equation}
    \overrightarrow{E_{rad}}(\vec{r},t) = \frac{q}{c}\left(\frac{\vec{n}}{\kappa^3 R} \times ((\vec{n} - \vec{\beta}) \times \dot{\vec{\beta}})\right),
\end{equation}
and
\begin{equation}
    \overrightarrow{B_{rad}}(\vec{r},t) = \vec{n} \times \overrightarrow{E_{rad}}.
\end{equation}
Here, $\vec{n} = \vec{R}/R$ and $\kappa = 1 - \vec{n}\cdot\vec{\beta}$. It is noted that, consistent with source free radiation solutions of Maxwell's equations, $|\vec{E_{rad}}| = |\vec{B_{rad}}|$. Constraining $\beta \approx 1$ (non- to semi-relativistic) simplifies the electric radiation field to 
\begin{equation}
    \overrightarrow{E_{rad}}(\vec{r},t) = \left(\frac{q}{Rc^2}\vec{n} \times (\vec{n} \times \dot{\vec{v}})\right).
\end{equation}
Taking the magnitudes of the fields gives
\begin{equation}
    |\overrightarrow{E_{rad}}| = \frac{q\dot{\vec{v}}}{Rc^2}sin(\Theta).
\end{equation}
This magnitude can be squared to recover the Poynting vector of the test particle
\begin{equation}
    S = \frac{c}{4\pi}|\overrightarrow{E_{rad}}|^2 = \frac{c}{4\pi}\left(\frac{q\dot{\vec{v}}}{Rc^2}sin(\Theta)\right)^2.
\end{equation}
Transforming the outward flow of energy to an energy per unit time $dW/dt$ per unit solid angle $\Omega$ (a more physically enlightening quantity) can be calculated by multiplying $S$ by $dA = R^2d\Omega$. Then, integrating over all solid angles recovers the power emitted per unit time $P$, or the Larmor Formula
\begin{equation}
    P = \frac{2q^2\dot{\vec{v}}}{3c^3}.
\end{equation}
where we may expand $\dot{\vec{v}} = \vec{a} = (\vec{a_{||}} + \vec{a_\perp})$ in terms of acceleration.\\

\indent With a basis set, we may now return to relativistic $\beta > 1$ particles and their radiation. Recalling  the equations of motion
\begin{equation}
    \frac{d\gamma m \vec{v}}{dt} = \frac{q}{c}\vec{v} \times \vec{B},
\end{equation}
and
\begin{equation}
    \frac{d\gamma m c^2}{dt} = q\vec{v} \cdot \vec{E},
\end{equation}
the following assumptions that $\gamma$ is constant used for this derivation. With this, we may allow
\begin{equation}
    m\gamma\frac{d\vec{v}}{dt} = \frac{q}{c}\vec{v} \times \vec{B}.
\end{equation}
We denote parallel $||$ and perpendicular $\perp$ to be with respect to the magnetic field. Therefore, $\vec{v_{||}} = 0$ and 
\begin{equation}
    \frac{d\vec{v_\perp}}{dt} = \frac{q}{\gamma m c}\vec{v_\perp} \times \vec{B}.
\end{equation}
From this, we may define a frequency of gyration as 
\begin{equation}\label{eq:gyro}
    \omega_g = \frac{qB}{\gamma\ m c},
\end{equation}
which will be used later in this section.\\

\indent Using the Larmor formula for relativistic particle emission 
\begin{equation}
    P = \frac{2q^2}{3c^2}\gamma^4(a_\perp^2 + \gamma a_{||}^2),
\end{equation}
where we eliminate $a_{||} = 0$. For relativistic emission, we define $a_\perp = \omega_g v_\perp$ and substitute into the Larmor formula
\begin{equation}
    P = \frac{2q^2}{3c^3}\gamma^4\frac{qBv_\perp}{\gamma m c}^2.
\end{equation}
Given a distribution of velocities, and not a single particle, the average with respect to $\beta = v/c$ over all possible angles of emission recovers the full synchrotron power total. Thus,
\begin{equation}
    <\beta_\perp> = \frac{\beta^2}{4\pi}\int sin^2(\alpha)d\Omega = \frac{2}{3}\beta^2,
\end{equation}
recovers
\begin{equation}
    P = \left(\frac{2}{3}\right)^2r_0^2c\beta^2\gamma^2B^2.
\end{equation}
Here, the classical electron radius $r_0 = e^2/mc^2$, magnetic energy density $U_B = B^2/8\pi$, and Thompson scattering cross section $\sigma_T = 8\pi r_0^2/3$.\\

\indent Throughout these derivations, we ignored the scale of the magnetic fields. Assuming that all magnetic fields produce synchrotron radiation given a relativistic particle is now wholly incorrect for the Weibel instability system.  Consider a relativistic particle moving through random small-scale magnetic fields. The term small-scale is used when the field's characteristic length $\lambda_B = 2\pi/k_B$ is less than or comparable to the electron's Larmor radius $r_e = \gamma\beta m c^2/eB_z$. We define $k_B$ as the correlation scale of the field
\begin{equation}
    k_B = \frac{4\gamma_{Beam}\omega_{p,e^-}}{2^{1/4}\bar{\gamma_e}^{-1/2}c},
\end{equation}
where $\gamma_{Beam}$ is the Lorentz factor of the shock and $\bar{\gamma_e}$ is the thermal Lorentz factor of the particles in the beam (known as delgam in simulations). The spectra produced by this particle depends on the ratio of its deflection or pitch angle $\alpha$ and its beaming angle $\Delta\theta$ \citep{LL2} . This can be estimated as 
\begin{equation}
    \frac{\alpha}{\Delta\theta} \sim \frac{eB_z\lambda_B}{mc^2} \sim \gamma\lambda_B r_e^{-1} \sim \frac{\gamma}{k_Br_e},
\end{equation}
which we now define as the Jitter Parameter $\delta_J \equiv \alpha/\Delta\theta$. \\

\indent It has been shown \citep{Brett2013, 5, 19, MVM2000} that values of $\delta_J$ can be binned into four regimes of interest. First, $\delta_J \rightarrow \infty$ corresponds toc classical synchrotron radiation (described above). Particles complete full orbits perpendicular to the direction of a homogeneous magnetic field. $\delta_J > \gamma$ occurs when a particle's Jitter parameter is greater than its Lorentz factor. The guiding center of the orbit drifts slowly due to an inhomogeneous magnetic field, creating a slanting helical trajectory crossing fields of varying strengths. While similar to pure synchrotron, this spectra is denoted as diffusive synchrotron radiation. Third,  the particle does not complete its full orbit due to short scale $B$-field variance. This produces $1 < \delta_J < \gamma$. This produces a pulsed emission with random intervals as seen by an observer. This spectra is two-fold. Near the spectral peak the radiation shows synchrotron in shape, but differs significantly at lower frequencies. This regime is commonly called the large-angle jitter radiation regime. Last, when $\delta < 1$ the deviations of the particles trajectory as smaller than its beaming angle. This results in patchy spectra seen by the observer as the particle produces emission over random patches of $B$-field. Known as small-angle jitter radiation, the resulting spectra is visibly different from synchrotron.  \\

\indent Despite the different regimes of spectra that can be recovered, they all produce the same total radiated power
\begin{equation}
    P_{tot} = \frac{2}{3}r_0^2c\gamma B_z^2.
\end{equation}

\subsection{Magnetic Fields}
\indent With the radiation sources and regimes of the WI system detailed, the method to disentangle them shifts to a different form of analysis. Radiative processes are known to be complex. Due to the turbulence present in the small scale field regime, studying the radiation produced by particles in these environments is just as complex\citep{Brett2013}. Therefore, we look to the fields generated by $e^\pm$ in the system to help elucidate the radiation mechanisms and sources. This summary comes from \citep{9, 19}.\\

\indent We begin be recalling in the WI, the magnetic field grows until it reaches saturation, where the field reaches equipartition to the kinetic energy of the particles. Mathematically, the fields generated by the $e^\pm$ via WI is
\begin{equation}
    B_e^2 = \eta_e \gamma_{Beam}mc^2n(8\pi)^{-1}.
\end{equation}
Here, $\eta_e$ is the efficiency factor for electrons. This ad hoc variable incorporates uncertainties due to the nonlinear phase of the WI and ranges from $0.1 - 0.01$, as seen from simulations. The jitter parameter for the electrons is also defined by
\begin{equation}
    \delta_e = 2^{-7/4}\sqrt{\frac{\bar{\gamma_e}}{\gamma_{init}}}
\end{equation}
This value can change as the WI evolves and therefore can be present in a variety of ranges (discussed above as $\delta_J$). Hereafter,  $\delta_J = \delta_e$ to avoid confusion. \\

\indent To fully uncover the jitter spectrum, the spatial scale of $B$ is required. This is difficult to obtain in first-principles due to the non-linearity of the instability. To rectify some of the difficulty, the Fourier transform of the field can be performed to recover some scale separation. This is done via power law
\begin{equation}\label{eq:Bk}
    B_k = C_Bk^\mu.
\end{equation}
This holds iff $0 \leq k \leq k_B$ and is $0$ otherwise. $\mu \geq 1$ is the spectral index of the power spectra. $C_B$ is a normalization constant defined using the total pulse duration of the radiation $T$
\begin{equation}
    C_B^2 = \pi(2\mu + 1)cT\bar{B_e^2}k_B^{-(2\mu + 1)},
\end{equation}
Defining the minimum scale as $k_{min} = \delta_ek_{B}$, a small-scale, sub-Larmor magnetic field can be formulated as
\begin{equation}
    \bar{B}_{SS}^2 = \int^{k_B}_{k_min} B_k^2 dk
\end{equation}
This small scale field has a ratio with the total electron generated fields 
\begin{equation}
    \frac{\bar{B}_{SS}^2}{\bar{B}_{e}^2} = 1 - \delta_J
\end{equation}
Given the magnetic field structure present in the system, the following can be deduced. There is a magnetic field $\bar{B}_{e}$ that is generated by the particles. A non-negligible fraction of this field is sub-Larmor in scale and described by (Eq. \ref{eq:Bk}) - power law with index $\mu$. A large scale magnetic field $\bar{B}_{LS} = \bar{B}_{e} - \bar{B}_{SS}$ is also present. These will be the fields associated with the current filamentation mechanisms. \\

\indent Radiation from $\bar{B}_{LS}$ is given by synchrotron radiation, while radiation by $\bar{B}_{SS}$ is now formally denoted as Jitter radiation. To recover a power expression, first recall the full expression for power per unit angle per unit frequency (found in \citep{RL1979, LL2})
\begin{equation}
    dW = \left(\frac{e^2\omega^4}{4\pi^2 c^3 \omega^{`4}}\right) \left|\vec{n} \times \left(\left(\vec{n} - \vec{\beta}\right) \times \overrightarrow{w_{\omega^`}}\right)\right|^2 d\Omega d\omega
\end{equation}
This can be simplified to spectral energy per unit frequency
\begin{equation}
    \frac{dW}{d\omega} = \frac{e^2\omega}{2\pi c^3}\int_{\omega/2\gamma^2} \frac{|\overrightarrow{w_{\omega^`}}|}{\omega^{`2}} \left(1 - \frac{\omega}{\omega^` \gamma^2} + \frac{\omega^2}{2\omega^{`2}\gamma^4}\right) d\omega^`
\end{equation}
where $\overrightarrow{w_{\omega^`}} = eB_k/\gamma mc$ is the Fourier component of the acceleration $a_\perp$.  Recalling the definition for gyrofrequency (Eq. \ref{eq:gyro}), we may replace the $B_z$ with $B_k$ to find the first term in parenthesis is a result of the jittering motion the electron experiences as it moves through the magnetic field turbulence. \\

\indent  Finally, the spectral power can be expressed by dividing $dW/d\omega$ by the total pulse duration $T$ \citep{RL1979}
\begin{equation}
    P(\omega) = r_0^2 c \gamma^2 \frac{\bar{B}_{SS}^2}{2\omega_J}J\left(\frac{\omega}{\omega_J}\right).
\end{equation}
$\omega_J$ is the characteristic frequency of the radiation $\omega_J = \gamma_2 k_B c$. Given the integral
\begin{equation}
    I(\xi) = \int \xi^{-2\mu}(1 - \xi + .5\xi^2)d\xi,
\end{equation}
the function $J$ is defined as 
\begin{equation}
    J(\xi) = (2\mu + 1)\xi^{2\mu}(I(min(2,\xi/\delta)) - I(\xi)).
\end{equation}
This allows us to simplify further to
\begin{equation}
     P(\omega) = \frac{e^2\delta^2\omega_J}{2c\gamma^2} \frac{\bar{B}_{SS}^2}{\bar{B}_{SS}^2}J\left(\frac{\omega}{\omega_J}\right).
\end{equation}
Integrating over all frequencies recovers
\begin{equation}
    \frac{dW}{dt} = \frac{2}{3}r_0^2 c \gamma^2 \bar{B}_{SS}^2,
\end{equation}
which is identical to synchrotron radiation when $\bar{B}_{SS}^2 = B_z$ (no small-scale fields).

\section{Results}
\subsection{Mathematical Equivalency}
\indent Before the simulation can be analyzed, self consistent units and dynamical equations must be confirmed. The relativistic motion equations (Eqs. \ref{Eq:2D_Energy_Eq_Final}, \ref{Eq:2D_x_Momentum_Eq_Final}, and \ref{Eq:2D_y_Momentum_Eq_Final}) are checked numerically for each snapshot for the entire particle population. These checks are crucial to ensuring the theory can be properly applied to the data, that the data, the theory, and the conclusions are all in the same units, and that the data is in a proper normalized state and in proper units. \\

\subsection{Heating}
\begin{figure}[h]
    \begin{subfigure}
        \centering
        \includegraphics[width=0.4\linewidth]{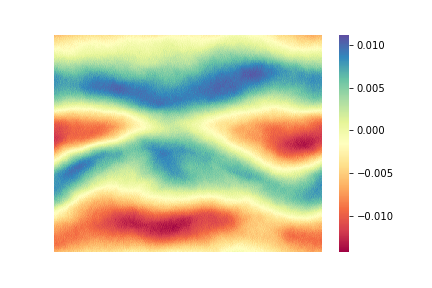}
    \end{subfigure}
    \begin{subfigure}
        \centering
        \includegraphics[width=0.4\linewidth]{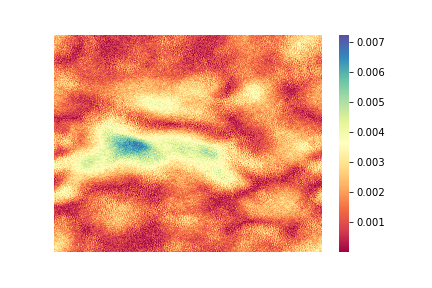}
    \end{subfigure}
    \begin{subfigure}
        \centering
        \includegraphics[width=0.4\linewidth]{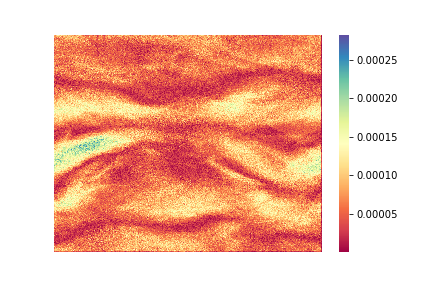}
    \end{subfigure}
    \begin{subfigure}
        \centering
        \includegraphics[width=0.4\linewidth]{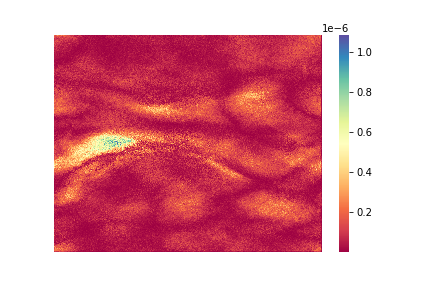}
    \end{subfigure}
    \caption{Field plots for $B_z$ (top left), $E$ (top right), $J$ (bottom left) and $E \cdot J$ (bottom right). The first and largest major filament merger occurs in the magnetic field and subsequent hot spots show up in the other fields. This heating hot spot is shown to be an area of particle acceleration.}
    \label{fig:JouleHotspots}
\end{figure}

\indent Within the 2D WI system, the source of heating will not come from pitch angle scattering as it does in 3D. Instead, it comes from a the Joule Heating product $\vec{E} \cdot \vec{J}$ (Fig. \ref{fig:JouleHotspots}). After the WI has saturated and begun to dissipate, filaments begin to slowly work toward coalescence as like-currents pull filaments together. As shown by the Poynting theorem of energy conservation, what energy does not get radiated out is transformed into heat as shown, or injected into the particles as a form of dynamical energy, thus accelerating them. \\

\subsection{Particle Energization}
\begin{figure}
    \begin{subfigure}
        \centering
        \includegraphics[width=0.4\linewidth]{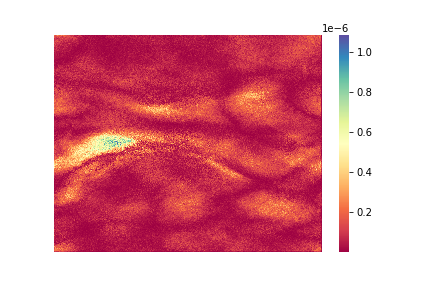}
    \end{subfigure}
    \begin{subfigure}
        \centering
        \includegraphics[width=0.4\linewidth]{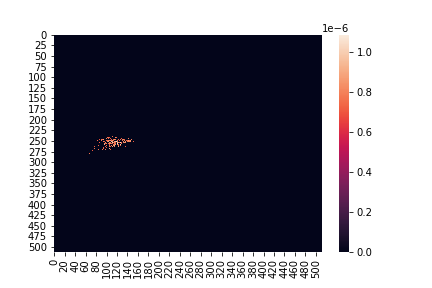}
    \end{subfigure}
    \begin{subfigure}
        \centering
        \includegraphics[width=0.4\linewidth]{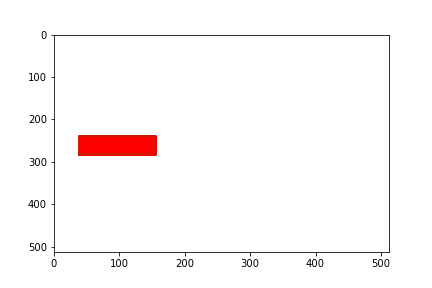}
    \end{subfigure}
    \begin{subfigure}
        \centering
        \includegraphics[width=0.4\linewidth]{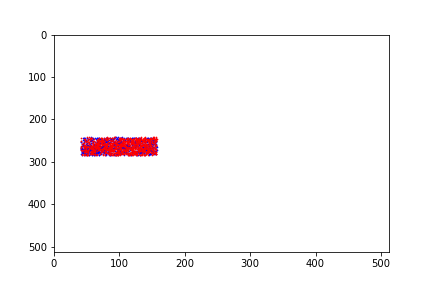}
    \end{subfigure}
    \caption{Various stages of isolation of the heating hot spot on the $512^2$ simulation grid (denoted by axis tick marks) . The full field is seen in the top left and a filtered region in the top right. The numerically isolated area can be seen in the bottom left. The particles that are located in these grid spots can be seen in the bottom right all converged within the box. }
    \label{fig:Merger}
\end{figure}

\begin{figure}
    \begin{subfigure}
        \centering
        \includegraphics[width=0.4\linewidth]{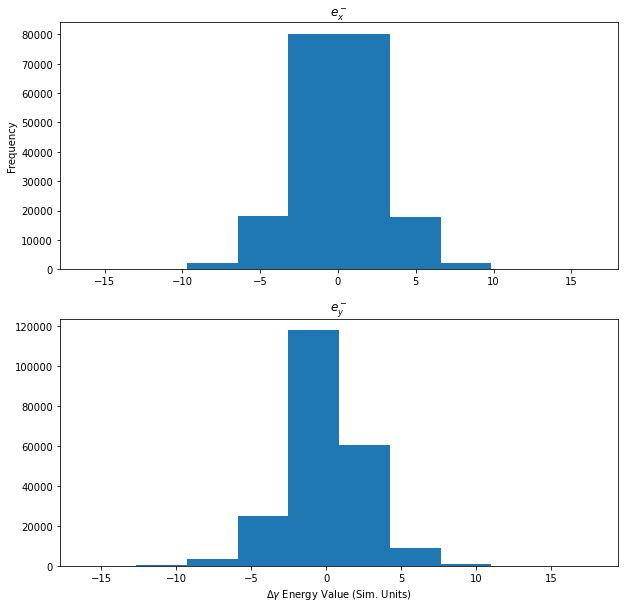}
    \end{subfigure}
    \begin{subfigure}
        \centering
        \includegraphics[width=0.4\linewidth]{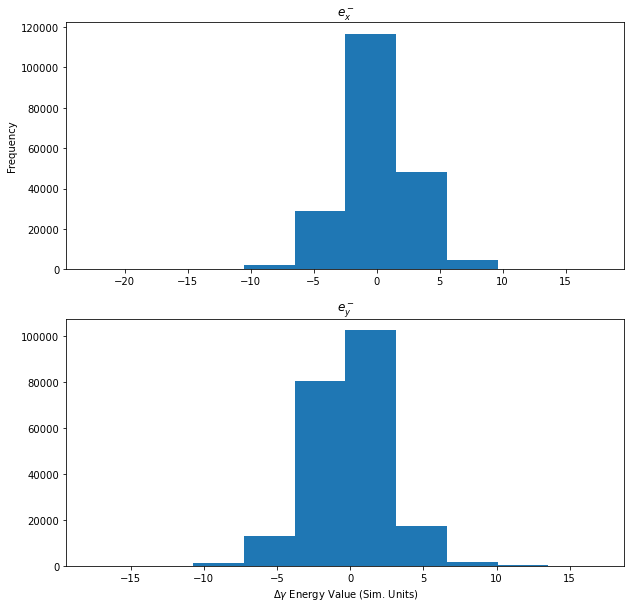}
    \end{subfigure}
    \begin{subfigure}
        \centering
        \includegraphics[width=0.4\linewidth]{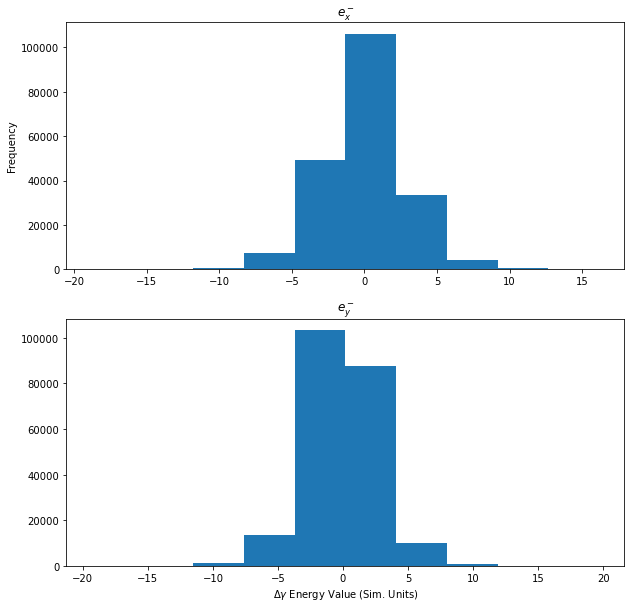}
    \end{subfigure}
    \begin{subfigure}
        \centering
        \includegraphics[width=0.4\linewidth]{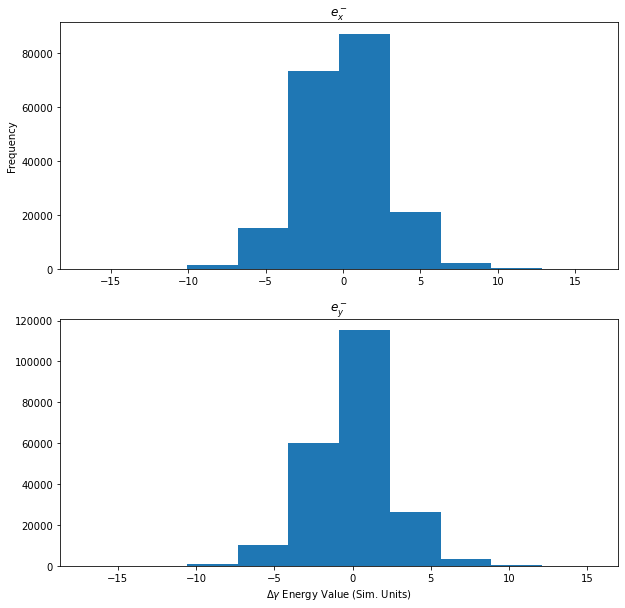}
    \end{subfigure}
    \caption{Histograms of $\Delta\gamma_{Energy}$ for the particles within the heating region. Electrons $e^-$ are shown on the top panel of the pairs while positrons $e^+$ are below. Any difference between the particle distributions is purely circumstantial and not based on physics. The snapshot that the isolation takes place is shown in the top right, with the snapshot before it to its left and the two after below. Positive values of $\Delta\gamma_{Energy}$ confirm particle energization and acceleration.}
    \label{fig:deltaGamma}
\end{figure}

\indent To show that particles indeed gain a form of dynamical energy, a singular filament merger is isolated (Fig. \ref{fig:Merger}). Using the simulation output for energy per particle $\gamma_{Energy} = \sqrt{1 + \vec{P}^2}$, the difference between values per snapshot is easily computed ($\Delta\gamma_{Energy}$) (Fig. \ref{fig:deltaGamma}). It is clear from the histograms that within the population of isolated particles, a fractional population gains energy at the snapshot in question, and subsequent snapshots. There is also a population of particles that looses energy during this heating process. When all other heating events are taken into account, this maintains the collective behavior and conservation of energy.

\clearpage

\subsection{Radiative Processes of Realistic Scenarios}
\begin{figure}[h!]
    \centering
    \includegraphics[width=0.5\linewidth]{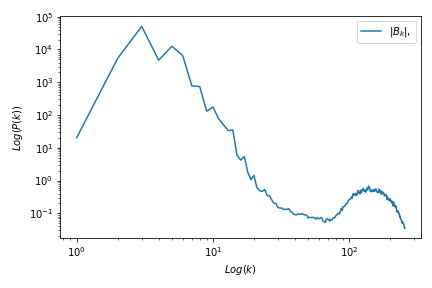}
    \caption{Spectral trend of the magnetic field in the Fourier regime as a function of wave number $\tilde{B}_z(k)$. The synchrotron signal can be seen at low wave numbers representing the growing current filaments. The numerical Cherenkov instability is seen at the very small physical scales (high $k$).}
    \label{fig:Bzk}
\end{figure}

\indent Radiative processes are known to be complex. It is no easy task to separate out various forms of radiation when more than one exist in a single system, and the WI environment is no exception. Untangling the synchrotron radiation and the jitter radiation in this current set-up is untenable. This work, and the current simulation in particular, is parameterized in a way to replicate GRBs in a more realistic and common setting, notable here with $\gamma_{Beam} = 3$. It is confirmed that jitter radiation must be in the system, coexisting with the synchrotron radiation. However, at the current beam propagation velocity, the system falls fully into the diffusive synchrotron radiation regime. The radiation peaks at larger physical scales and is synchrotron in nature, but becomes convoluted at larger wavenumbers (small physical scales). This is where jitter radiation is expected to be seen. Viewing $B_z(k)$, the picture becomes clearer (Fig. \ref{fig:Bzk}). Taken while filament building is happening, the known synchrotron shape is immediately seen at low-$k$ (corresponding to the growing filaments). Next, the semi-circular mode on the tail of the trend is the numerical Chenrenkov instability. This is known to be present and serves here as a good indication that the physics in the simulation are accurate. \\

\begin{figure}[h!]
    \centering
    \includegraphics[width=0.6\linewidth]{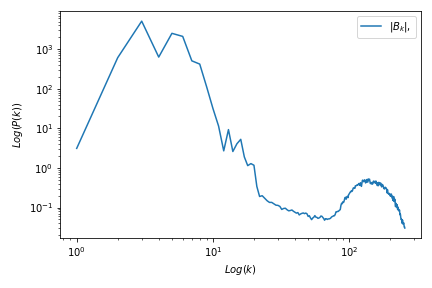}
    \caption{A clear view of all three modes seen in the scale separation of the magnetic field. The middle regime is where the jitter radiation is found. This is the diffusive synchrotron signal.}
    \label{fig:BkBump}
\end{figure}

\begin{figure}
    \centering
    \includegraphics[width=0.8\linewidth]{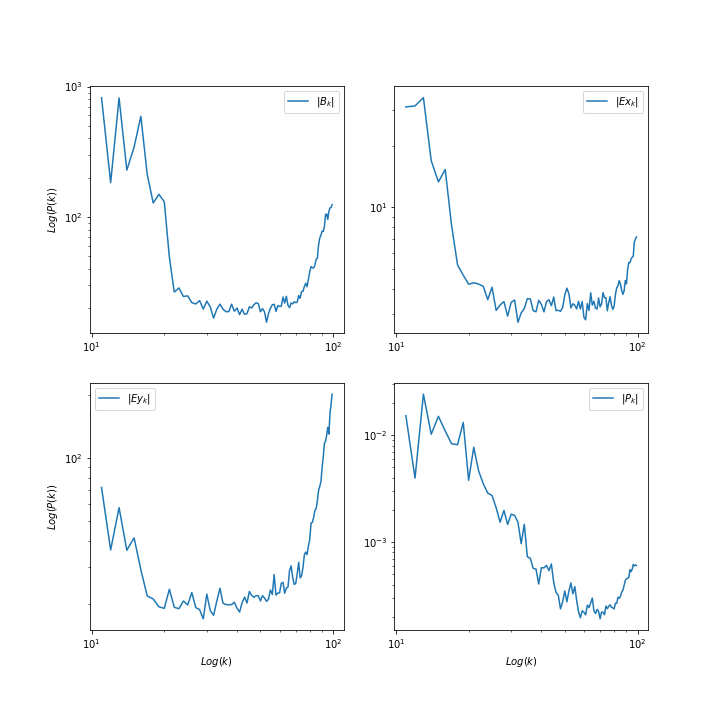}
    \caption{The radiation from the diffusive synchrotron region for $B_z$ (top left), $E_x$ (top right), $E_y$ (bottom left), $P$ (bottom right). It is seen that there are no discernible trends or behaviors.}
    \label{fig:BumpFields}
\end{figure}

\indent Next, the behavior in the middle of the trend, seen in (Fig. \ref{fig:BkBump}) at a slightly earlier time in the simulation, is isolated. The radiation from the fields ($B_z$, $E_x$, $E_y$, $P$) is shown in (Fig. \ref{fig:BumpFields}). It is immediately clear there is no discernible pattern or shape present. This is the first indication that more realistic scenarios are more complex in nature.\\

\begin{figure}
    \begin{subfigure}
        \centering
        \includegraphics[width=0.4\linewidth]{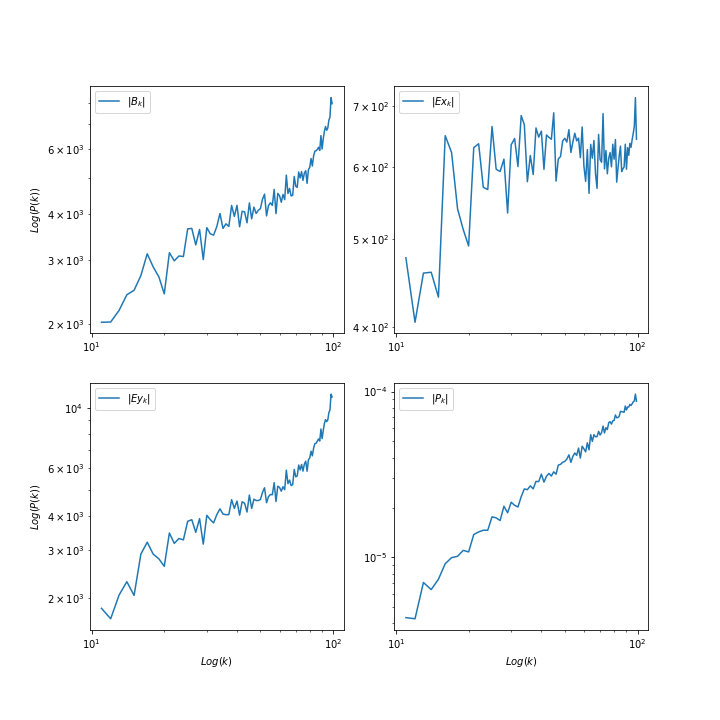}
    \end{subfigure}
    \begin{subfigure}
        \centering
        \includegraphics[width=0.4\linewidth]{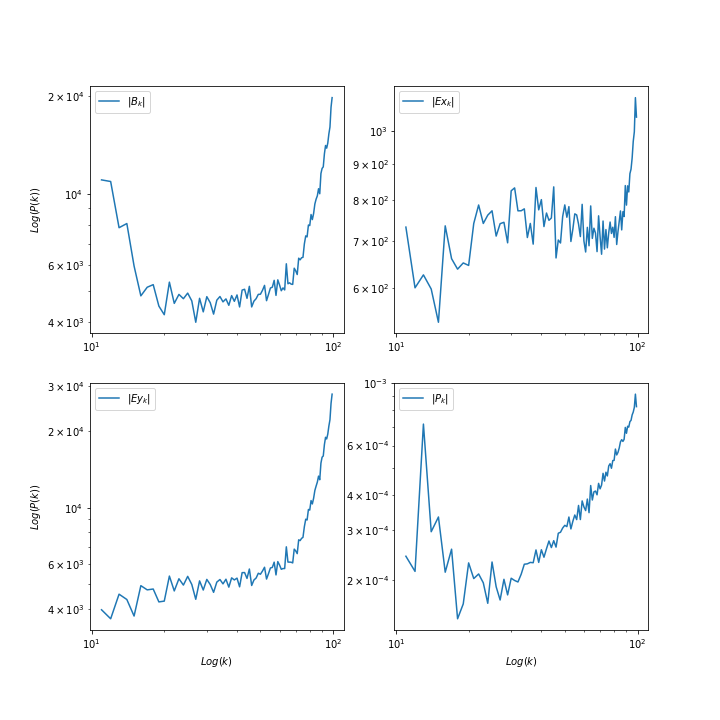}
    \end{subfigure}
    \begin{subfigure}
        \centering
        \includegraphics[width=0.4\linewidth]{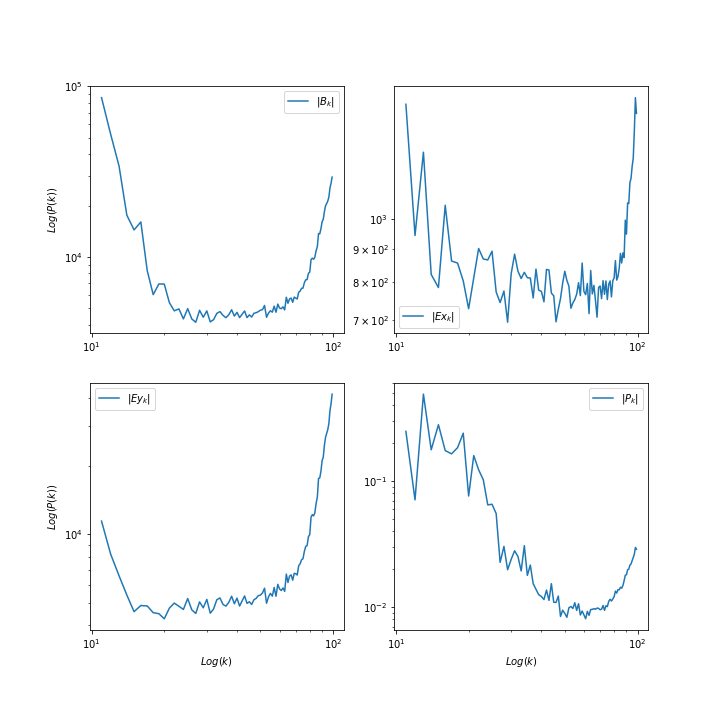}
    \end{subfigure}
    \begin{subfigure}
        \centering
        \includegraphics[width=0.4\linewidth]{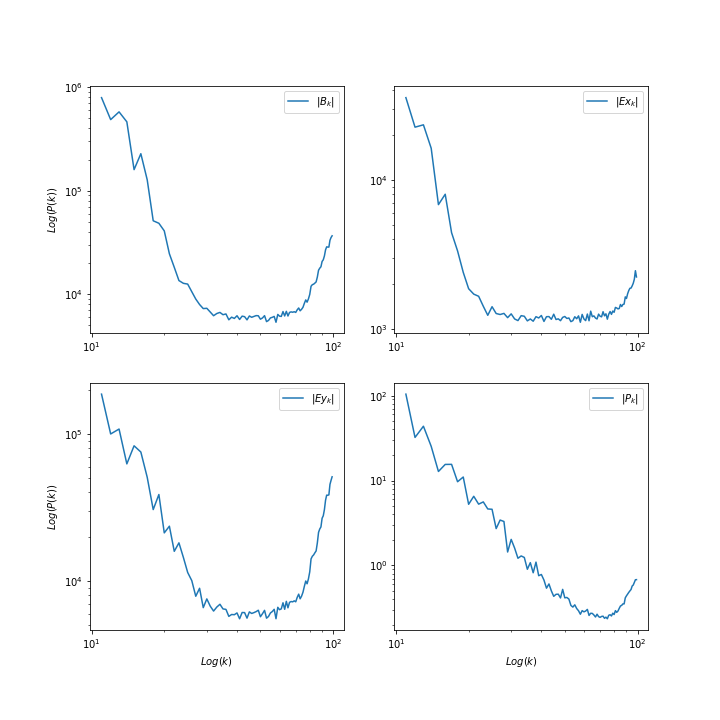}
    \end{subfigure}
    \caption{Trends for the first five epochs as a difference of radiated power between two epochs. Epoch $2 - 1$ is shown in the top left, $3 - 2$ in the top right, $4 - 3$ in the bottom left, and $5 - 4$ in the bottom right for $B_z$ (top left panels), $E_x$ (top right panels), $E_y$ (bottom left panels), $P$ (bottom right panels). The time prior to saturation (bottom right), where the WI is building filaments, will be when jitter radiation is at its max. It is clear that the complex nature of radiative processes makes it almost impossible to disentangle the two different radiation signals.}
    \label{fig:EpochTrends}
\end{figure}

\indent The last evidence of an overly complex system can be seen in the difference of the sum of all bins within an epoch. Taking the sum of each bin in an epoch and subtracting the value from the epoch before it presents information on how the whole of radiation changes between the self-defined epochs. Viewing the first four trends (epoch $2$ - epoch $1$, epoch $3$ - epoch $2$, etc.) confirms the findings (Fig. \ref{fig:EpochTrends}). Without any salient pattern or comparison to known theory, it must be concluded that while present, the nature of jitter + synchrotron radiation is too complex at this scale to disentangle. Jitter radiation is known to exist in Weibel mediated shock systems and is most prevalent prior to saturation, where magnetic field turbulence from growing current filaments can be at its most complex. This work looked at a more common set up found in nature, specifically, in reference to the Lorentz factor of the shock (or counter streaming beams). A value of $\gamma_{Beam} = 3$, while realistic in nature, is far too slow for the separation of scales needed to see jitter radiation outright. The current set-up falls firmly into the diffusive synchrotron regime. A clear cut, visual confirmation of separate of regimes (small angle jitter) requires ultra-relativistic flows ($\gamma_{Beam} > 10$). This scenarios in nature would be much more rare than the system simulated here. This would also require changes to other parameters as well (ex. skin depth, particle density, and box size).

\onecolumngrid

\clearpage
\newpage
\twocolumngrid

\section{Conclusions}

\indent The WI system is visualized with massive current-filaments created by the instability itself. It has been shown through current filament studies \citep{20} and by full system parameterizations \citep{MCS_WP1} that these filaments interact and dictate the next phase of the system, the Two Stream instability. Merging of filaments has been shown to generate energy \citep{MCS_WP1} and radiative power \citep{MCS_WP2} and are a possible host for collisionless heating. Filaments can be visible in the magnetic fields, but they are mainly described as current carrying structures. And while the complex physics of filament coalescence can be safely ignored, the possible outcome of this energy dissipation/convergence can not. This paper detailed the existence and effect of collisionless heating at post-saturation in the Weibel instability system. Because the system exists solely in 2D, the common mechanism for heating, pitch angle diffusion, can not be the cause. This is due to a constant pitch angle of $\pi/2$. Heating is therefore a byproduct of the merging current filaments directly through the conservation of energy. What little energy isn't transferred into the particles, seen in the spectral analysis, is transformed into heat. The newly energized particles are shown to accelerate. Confirming that heating is a valid mechanism for particle energrization even within non-collisional plasmas. \\

\indent Violent filament merger events could serve as the accelerators for particles to breach thermal velocities and be active participants in diffusive shock acceleration. Parameterization studies show that given favorable but not-unlikely conditions, Weibel dissipation is interrupted by these violent merger events within the 2D counter-streaming setup. The violent mergers are shown to interact with spectral waves and produce a secondary excited state in the Two Stream instability. These mergers have also be shown to heat the local plasma and accelerate the local particle population. These accelerations produce radiation not unlike those seen in diffusive synchrotron regimes. The Weibel mediated multi-instability system brings forth favorable means to produce a solution to the injection problem in collisionless pair-plasma shocks.



%
%

%

\begin{acknowledgments}
This study is partly supported by PHY-2010109. The author would like to thank Dr. Medvedev for his guidance and role of graduate advisor for this work and subsequent works. This research made use of "Tristan-MP v2" particle-in-cell code.
\end{acknowledgments}

\appendix
\nocite{*}
\bibliography{main}
\end{document}